\documentstyle[aps,twocolumn]{revtex}

\newcommand{\unit}{\hbox{$\hat{\bf n}$}}

\newcommand{\rv}{{\bf r}}
\newcommand{\Ev}{{\bf E}}

\newcommand{\dv}{{\bf d}}

\newcommand{\kappav}{\bbox{\kappa}}
\newcommand{\Dkappav}{\Delta\bbox{\kappa}}

\newcommand{\kv}{{\bf k}}
\newcommand{\eo}{\epsilon_0}
\newcommand{\beq}{\begin{equation}}
\newcommand{\eeq}{\end{equation}}
\newcommand{\bea}{\begin{eqnarray}}
\newcommand{\eea}{\end{eqnarray}}

\begin{document}
\draft
\title{Pumping two dilute gas Bose-Einstein condensates \\with Raman
light scattering}
\author{ C.M. Savage$^{1,2}$, Janne Ruostekoski$^{1}$ and Dan
F. Walls$^{1}$}
\address{$^{1}$Department of Physics, University of Auckland, Private Bag
92019, \\
Auckland,New Zealand \\
$^{2}$Department of Physics
and Theoretical Physics, Faculty of Science, Australian National
University, ACT 0200, Australia}

\date{\today}
\maketitle
\begin{abstract}
We propose an optical method for increasing the number of atoms in a
pair of dilute gas Bose-Einstein condensates.  The method uses
laser-driven Raman transitions which scatter atoms between the condensate
and non-condensate atom fractions.  For a range of condensate phase
differences there is destructive quantum interference of the
amplitudes for scattering atoms out of the condensates.  Because the
total atom scattering rate into the condensates is unaffected the
condensates grow.  This mechanism is analogous to that responsible for
optical lasing without inversion.  Growth using macroscopic quantum
interference may find application as a pump for an atom laser.
\end{abstract}
% 03.75.Fi : Bose condensation
% 05.30.Jp : Boson systems
% 42.50.Vk : Mechanical effects of light on atoms
% 32.80.Pj : Optical cooling of atoms, trapping

In the recent experiments demonstrating Bose-Einstein condensation of
alkali vapors the first stages of cooling are optical \cite{becexp}.
The final stage utilises evaporation of the hottest atoms out of the
trap \cite{HES87}.  Despite the great success of evaporation it has
the disadvantage of removing atoms from the system.  Consequently
alternative final stage cooling methods are being investigated.
Velocity selective coherent population trapping (VSCPT) is one optical
method potentially capable of cooling to the Bose-Einstein transition
point \cite{vscpt}.

In this paper we analyze an optical method for increasing the number
of atoms in a pair of overlapping Bose-Einstein Condensates (BECs),
such as have been produced in the laboratory using sympathetic cooling
\cite{Myatt,Julienne}.  According to conventional spontaneous symmetry
breaking arguments BECs are in coherent states with a definite global
phase \cite{FOR75}.  A pair of BECs therefore has a definite phase
difference, which can be measured by a variety of techniques
\cite{JAV96b,IMA96,Ruostekoski96,Savage97,Ruostekoski97}.  We utilise
this phase difference to suppress transitions of atoms out of the
condensates while driving transitions into the condensates, producing
a net condensate growth.  The transitions are driven by spontaneous
Raman scattering of laser light \cite{Ruostekoski96}.  For a certain
range of condensate phase differences transitions out of the
condensates are suppressed by destructive quantum interference.  The
interference is between the amplitudes for transitions from each
condensate into the non-condensate atom fraction.

This suppression of atom scattering out of the condensates is analogous
to the suppression of absorption that occurs in optical lasing without
inversion \cite{LWI}.  In both cases the suppression is due to
destructive quantum interference between two channels to the same
final state.  This mechanism for condensate growth
might also be suitable as an optically driven pump for an atom laser
\cite{atomlaser}.

Spontaneous Raman scattering of laser light from two BECs has been 
analyzed by Ruostekoski and Walls \cite{Ruostekoski96}, whose analysis 
we shall follow.  They showed that the scattered light spectrum 
depends on the phase difference between the two condensates.  The 
spectrum has two peaks, corresponding to transitions of atoms into and 
out of the condensates.  For certain values of the condensate phase 
difference the second peak disappears due to the destructive quantum 
interference previously described.  Although this suggests the 
possibility of condensate growth, only the short time behaviour was 
considered.  In fact, the Raman lasers induce Josephson oscillations 
between the condensates \cite{JAV86}.  We show that growth occurs and 
can persist over a complete period of this dynamics.  However the 
necessary destructive interference only occurs for a particular range 
of condensate phase differences.  Consequently, condensate growth only 
occurs for a subensemble of condensate pairs.  Suitable subensemble 
members might be chosen after a measurement of the phase difference 
\cite{JAV96b,IMA96,Ruostekoski96,Savage97,Ruostekoski97}.

We consider the set-up first introduced by Javanainen {\cite{JAV96b}}; 
a quantum degenerate gas with two ground states and a common excited 
state.  All the ground state atoms are confined in the same trap.  The 
pair of condensates are in two different Zeeman sublevels 
$|b\rangle=|g,m\rangle$ and $|c\rangle=|g,m-2\rangle$.  The state 
$|c\rangle$ is optically coupled to the electronically excited state 
$|e\rangle=|e,m-1\rangle$ by the field $\Ev_{2}$ having a polarization 
$\sigma_+$ and frequency $\Omega_{2}$.  Similarly, the state 
$|b\rangle$ is coupled to $|e\rangle$ by the field $\Ev_{1}$ with a 
polarization $\sigma_-$ and frequency $\Omega_{1}$.  Following Ref.  
\cite{JAV95b} the Hamiltonian density for the system is 
\bea
{\cal H}&=&
\psi^\dagger_{b} H \psi_{b}+
\psi^\dagger_{c} ( H +\hbar\omega_{cb}) \psi_{c}
\nonumber\\
&&+ \psi^\dagger_{e} ( H+\hbar\omega_{eb}) \psi_{e}+
{\cal{H}}_F
\nonumber\\
&&- \left(\dv_{b}\cdot\Ev_1 \,\psi^\dagger_{b} \psi_{e} +
\dv_{c}\cdot\Ev_2 \,
\psi^\dagger_{c} \psi_{e} +{\rm h.c.}
\right)\, .
\label{eq:HDN}
\eea
The first three terms reflect the center of mass energy, $H$, and the
internal energies of the atoms in the absence of electromagnetic
fields.  The frequencies for the optical transitions $e\leftrightarrow
b$ and $e\leftrightarrow c$ are $\omega_{eb}$ and $\omega_{ec}$
($\omega_{cb}=\omega_{eb}-\omega_{ec}$), respectively.  ${\cal H}_F$
is the Hamiltonian density for the free electromagnetic field.  The
final, bracketed, terms are for the atom-light dipole interaction.
The dipole matrix element for the atomic transition $e\leftrightarrow
b$ ($e\leftrightarrow c$) is given by $\dv_{b}$ ($\dv_{c}$).

We assume that the driving light fields ${\bf E}^+_{di}$ are in
coherent states and detuned far from single photon resonance so that
multiple scattering can be ignored.  We also assume they are plane
waves propagating in the positive $z$ direction with wavevectors
$\kappav_{i}$
\begin{mathletters}
\bea
\tilde{\bf E}^+_{d1}({\bf r}) &=& {1\over2}
{\cal E}_{1} \hat{\bf e}_{-}
\exp( i \kappav_{1} \cdot {\bf r} )\, ,\\
\tilde{\bf E}^+_{d2}({\bf r}) &=& {1\over2}
{\cal E}_{2} \hat{\bf e}_{+}
\exp( i \kappav_{2} \cdot {\bf r} )\, ,
\label{eq:INF}
\eea
\end{mathletters}
where $\hat{\bf e}_{i}$ are the unit circular polarization vectors, and we
have defined slowly varying fields by $\tilde{\bf E}^+_{i} = e^{i\Omega _{i}
t}{\bf E}^+_{i}$. We also define the slowly varying matter field
$\tilde{\psi_{c}} = e^{i (\Omega_{1}-\Omega_{2}) t} \psi_{c}$.

In the limit of large detuning the excited state field operator
$\psi_e$ may be eliminated adiabatically.  Following Javanainen and
Ruostekoski \cite{JAV95b} the scattered electric fields may then be
expressed in terms of the driving fields as $\tilde{\bf E}^{+}_{s} =
\tilde{\bf E}^{+}_{s1} +\tilde{\bf E}^{+}_{s2}$ where $\tilde{\bf
E}^{+}_{s1}$ is radiated by decays into state $|b \rangle$,
\bea
\tilde{\bf E}^+_{s1}({\bf r},t)
&=& \int d^3r' \,
{\bf K}({\bf d}_{b})
\psi^\dagger_{b} \psi_{e}
\nonumber \\
&=& {1\over\hbar\Delta_1} \int d^3r' \,
{\bf K}({\bf d}_{b})
\nonumber \\
&& \times
\left\{ \dv_{b} \cdot
\tilde{\bf E}^+_{d1} \, \psi^\dagger_{b} \psi_{b}
+\dv_{c} \cdot \tilde{\bf E}^+_{d2}
\,\psi^\dagger_{b}  \tilde{\psi}_{c}  \right\} \,.
\label{eq:FEQa}
\eea
The driving fields and atom fields are all functions of $\rv'$ and 
$t$.  $\Delta_{1} =\Omega_{1} -\omega_{eb}$ is the atom-field detuning 
of field 1.  The first line represents the radiation from the atomic 
dipole density, and the second follows after adiabatic elimination of 
the excited state.  $\tilde{\bf E}^{+}_{s2}$, which is radiated by 
decays into state $|c \rangle$, is found by swapping subscripts $b$ 
and $c$ and swapping the driving fields $\tilde{\bf E}^+_{d1}$ and 
$\tilde{\bf E}^+_{d2}$.  We have used the first Born approximation 
based on the assumption that the incoming fields dominate inside the 
sample, as multiple scattering is negligible.  The kernel ${\bf 
K}({\bf d})$ is the familiar expression \cite{JAC75} for the 
positive-frequency component of the electric field at $\rv$ from a 
monochromatic dipole with the complex amplitude ${\bf d}$, located at 
$\rv'$.

After adiabatic elimination of the excited state from the Hamiltonian
density Eq.~{(\ref{eq:HDN})}, and approximation of the electric fields by
the driving fields, the following Hamiltonian density is found to first
order in the inverse atom-field detuning \cite{Ruostekoski96}
\bea
&& {\cal H}_M = \psi^\dagger_{b}
( H -\hbar\delta_1 )
\psi_{b}+ \tilde{\psi}^\dagger_{c} ( H-
\hbar\delta_{cb}-\hbar\delta_2) \tilde{\psi}_{c}
\nonumber \\
&& +\hbar\kappa \left( \psi^\dagger_{b}
\tilde{\psi}_{c} \exp(-i \kappav_{12} \cdot \rv)+
\tilde{\psi}^\dagger_{c} \psi_{b} \exp(i \kappav_{12} \cdot \rv)
\right)\, ,
\label{eq:HDN3}
\eea
where $\kappav_{12} = \kappav_{1} -\kappav_{2}$ is the wavevector
difference of the driving light fields.  We have introduced the light-induced
level shifts $\delta_{i}$, the detuning from two-photon resonance
$\delta_{cb} = \Omega_{1} -\Omega_{2} -\omega_{cb}$, and the Raman coupling
coefficient $\kappa$
\beq
\delta_1={|{\cal E}_1|^2 d_{b}^2\over 4\hbar^2\Delta_1},\quad
\delta_2={|{\cal E}_2|^2 d_{c}^2\over 4\hbar^2\Delta_1},\quad \kappa=
{{\cal E}^*_1{\cal E}_2 d_{b}d_{c}\over 4\hbar^2\Delta_1} \, .
\label{eq:para}
\eeq
The dipole matrix elements $d_{b}$ and $d_{c}$ contain the reduced
dipole matrix elements and the corresponding nonvanishing
Clebsch-Gordan coefficients.  To simplify the algebra, we assume
$\kappa$ to be real.

The intensity of the scattered light at position $\rv$ is given by
\beq
I(\rv) = 2c \eo \langle
\tilde{\bf E}^{-}_{s} \cdot \tilde{\bf E}^{+}_{s} \rangle \, .
\label{intensity}
\eeq
Substituting in the expressions for the scattered fields in terms of
the atom fields, Eq.~{(\ref{eq:FEQa})}, generates a sum of terms for the
intensity of the form
\bea
2c \eo && \left( \frac{1}{\hbar \Delta_{1}} \right)^{2}
\int d^{3}r' d^{3}r''
[{\bf K}({\bf d}_{b})' ]^* \cdot {\bf K}({\bf d}_{b})''
 \nonumber \\
&& \times (\dv_{b}^{*} \cdot \tilde{\Ev}_{d1}^{-}) (\dv_{b}
\cdot \tilde{\Ev}_{d1}^{+})
\langle {\psi^\dagger_{b}}' {\psi_{b}}'
{\psi^\dagger_{b}}'' {\psi_{b}}'' \rangle \, .
\label{example term}
\eea
The $'$ and $''$ respectively denote functional dependence on $\rv'$
and $\rv''$.  We now assume that the driving fields have the same
wavevectors, so that $\kappav_{12}={\bf 0}$.  The dynamics of the
ground state fields $\tilde{\psi}_c$ and $\psi_b$ follows from the
Hamiltonian Eq.~(\ref{eq:HDN3}).  We assume a translationally
invariant and non-interacting Bose gas.  The matter field operators
are given by the familiar plane wave representations $\psi_b (\rv
t)=V^{-1/2}\sum_{\kv} \exp( i\kv \cdot {\bf r} ) \, b_{\kv}(t)$ and
$\tilde{\psi}_c (\rv t)=V^{-1/2}\sum_{\kv} \exp( i\kv \cdot\hbox{\bf
r} ) \, \tilde{c}_{\kv}(t)$, where $V$ is the mode volume.  In the
absence of light, the center of mass motion in both ground states
satisfies the dispersion relation $\epsilon_{\kv}=\hbar|\kv|^2/2m$,
with $m$ the atomic mass.  Defining the effective two-photon detuning
$2\bar{\delta}=\delta_{cb}-\delta_1+\delta_2$ and the condensate
oscillation frequency $\Omega_R=(\bar{\delta}^2+\kappa^2)^{1/2}$, the
mode operators at time $t$ are given in terms of the operators at time
$t=0$ by
\begin{mathletters}
\bea
\tilde{c}_{\kv}(t) &=& e^{i \alpha t}
\left\{
A \tilde{c}_{\kv}(0) - B b_{\kv}(0)  \right\} \,,
\\
b_{\kv}(t) &=& e^{i \alpha t}
\left\{
A^{*} b_{\kv}(0)  -B \tilde{c}_{\kv}(0) \right\} \, ,
\\
\alpha &=& \bar{\delta}+ \delta_1 -\epsilon_{ \kv } \, ,
\\
A &=& \cos{\Omega_R t} +{i\bar{\delta}\over \Omega_R} \sin{\Omega_R
t} , \,
B = i {\kappa\over\Omega_R} \sin{\Omega_R t} \, .
\label{define AB}
\eea
\label{eq:osc}
\end{mathletters}
Before the light is switched on, the atoms in the states $|b\rangle$
and $|c\rangle$ are assumed to be uncorrelated.  The Raman fields
induce a coupling between the two levels.  In the presence of Bose
condensates in the ground states, the coupling between the two
condensates is analogous to the coherent tunneling of Cooper pairs in
a Josephson junction \cite{JAV96b,JAV86}.
The expectation values of products of four atom field operators, such
as occurs in Eq.~{(\ref{example term})}, may be evaluated after
substituting in the expressions Eq.~{(\ref{eq:osc})} for the fields at
time $t$ in terms of the time zero fields.  For example the
expectation value in Eq.~{(\ref{example term})} becomes
\bea
&&\langle {\psi^\dagger_{b}}' {\psi_{b}}'
{\psi^\dagger_{b}}'' {\psi_{b}}'' \rangle |_{t} =
\nonumber \\
&& \langle \,
[ A {\psi^\dagger_{b}}' +B {\psi^\dagger_{c}}' ]
[ A^{*} {\psi_{b}}' -B {\psi_{c}}' ]
\nonumber \\
&& \times [ A {\psi^\dagger_{b}}'' +B {\psi^\dagger_{c}}'' ]
[ A^{*} {\psi_{b}}'' -B {\psi_{c}}'' ]
\, \rangle  \, ,
\label{example tzero}
\eea
where all the field operators on the right hand side are evaluated at
time zero.

The field operators are sums over the condensate and
non-condensate modes.  Since we are {\em only} interested in the
change in the number of condensate atoms due to light scattering we
need {\em only} evaluate those terms corresponding to scattering of atoms
into or out of the condensate, {\it i.e.} the incoherent part of the
scattering.  We ignore scattering of atoms between
non-condensate modes.  Together with momentum conservation this leads
to a considerable simplification of the terms like Eq.~{(\ref{example
tzero})}.  Once a particular plane wave mode is chosen for the first factor in
Eq.~{(\ref{example tzero})} the requirement for a non-zero expectation
value determines the modes occurring in all the remaining factors.
For example, the part of Eq.~{(\ref{example tzero})} relevant to
condensate depletion and growth is
\beq
\langle \,
D_{0}^{\dagger} D_{-} D_{-}^{\dagger} D_{0} +
D_{+}^{\dagger} D_{0} D_{0}^{\dagger} D_{+}
\, \rangle  \, ,
\label{example modes}
\eeq
where $D_{i}=A^{*}(t) b_{i}(0) -B(t) c_{i}(0)$ and the subscripts $0$
and $\pm$ respectively refer to the condensate mode and the
non-condensate modes having momenta $\pm \hbar \Dkappav$. Here
$\Dkappav = \Omega \unit/c-\kappav$ is the wavevector change of the
scattered photon, and $\unit = \rv/|\rv|$ is the unit vector in the
light scattering direction under consideration.  The two
non-condensate modes $+/-$ respectively arise from scattering of atoms
into/out of the condensate.  Note that the particular atomic mode denoted
depends on the light scattering direction $\unit$,  as does the
polarization of the scattered light.
In general, the polarizations of the emitted photons from the two
different atomic transitions are not orthogonal.  However, although
the resulting interference terms are nonvanishing in a particular direction
\cite{Ruostekoski96}, their contribution to the total intensity of the
scattered light vanishes after integration over all scattering
directions.  Because we are ultimately interested in the total
intensity, we ignore the terms proportional to $\langle \tilde{\bf
E}^{-}_{s1} \cdot \tilde{\bf E}^{+}_{s2} \rangle + {\rm c.c.}$ in
the intensity of the scattered light.

For brevity we assume that the number of atoms in the ground 
non-condensate states are the same, $n_{\pm} = \langle 
b^{\dagger}_{\pm} b_{\pm} \rangle = \langle c^{\dagger}_{\pm} c_{\pm} 
\rangle$, and that $n_{+}=n_{-}=n$ due to isotropy.  Further 
simplification occurs if we assume that there are equal numbers of 
atoms $N$ in each condensate, and that the laser intensities are 
chosen so that the level shifts are equal $\delta =\delta_{1} 
=\delta_{2}$.  Evaluating all the relevant terms in 
Eq.~{(\ref{intensity})} we find the following expressions for the 
intensity due to scattering of atoms into and out of the condensates 
\bea
I_{\rm in} &=& 2 C n \delta d_{s}^{2} \{
N + 2 {\rm Re}[ A^{*} B \langle c_{0}^{\dagger} b_{0} \rangle ]
d_{d}^{2} \} \, ,
\label{intensity in reduced}\\
I_{\rm out} &=& 2 C  ( n+1 ) \delta d_{s}^{2} \{
N +
{\rm Re}[ (A^{*2} -B^{2}) \langle c_{0}^{\dagger} b_{0} \rangle ] \}
\, ,
\label{intensity out reduced}
\eea
\beq
C = \frac{c_{L} |\kappav|^{4} }{8 \pi^{2} \eo\Delta_1}
\frac{1}{|\rv|^{2}} ( 1 -{1\over 2}\sin^{2} \theta) ,
\eeq
where $c_{L}$ is the speed of light, $d_{s}^{2} = d_{b}^2 +d_{c}^2$,
and $d_{d}^{2} =(d_{b}^2 -d_{c}^2)/d_{s}^{2}$.  We next integrate
these intensities over the Josephson oscillation period $P=2\pi /
\Omega_{R}$.  We find the time averaged intensities
\bea
\frac{1}{P}
\int_{0}^{P} I_{\rm in} \, dt &=& C' n \left\{ 1 +
\frac{ \bar{\delta} \kappa }{\Omega_{R}^{2}}
d_{d}^{2} \cos \Theta
\right\}
\, , \label{time1} \\
\frac{1}{P}
\int_{0}^{P} I_{\rm out} \, dt &=& C' ( n+1 ) \left\{ 1 +
\frac{\kappa^{2}}{\Omega_{R}^{2}} \cos \Theta
\right\}
\, , \label{time2}
\eea
where $C' = 2C \delta d_{s}^{2} N$, and $\Theta$ is the condensate 
phase difference.  The non-condensate populations $n$ are functions of 
$\Dkappav$ and hence of the scattering angle $\theta$ between $\unit$ 
and the laser propagation direction.  Our final step is integration 
over all scattering directions. This yields the total scattered light 
intensity and hence the total atom transition rates.  The angular 
integration has the effect of replacing $n$ by 
\beq
{8\pi\over3}\tilde{n} \equiv 2\pi \int_{0}^{\pi}
(1 -{1\over2} \sin^{2} \theta ) n(\theta)
\sin \theta \, d \theta \, ,
 \label{atompop}
\eeq
and $n+1$ by $8 \pi(\tilde{n} +1) /3$.  This integral may be
interpreted as the number of non-condensate atoms available for
scattering into the condensates.  In an infinite homogeneous system
the integral is divergent at the low energy end.  However, for a
finite, trapped system a low energy cutoff is provided by the first
excited state. A numerical integration assuming the
Bose-Einstein distribution at $T=400$ nK, and a low
energy cutoff of $\hbar (2\pi \times 100)$ J, gives
$(8\pi /3) \tilde{n} \approx 65$ for rubidium.  However, this is a
crude estimate since realistic systems are not expected to be in
thermal equilibrium.

Assuming, for simplicity, equal dipole moments $d_{d}=0$ the net
rate of scattering of atoms into the condensates (atoms per second) is
then, from Eqs.~(\ref{time1}-\ref{atompop}),
\bea
R &=& -6 \pi \left( \frac{\gamma}{\Delta_{1}} \right)^{2}
\left( \frac{I_{d}}{ \hbar c_{L} |\kappav|^{3}} \right) N
\nonumber \\
&& \times \left\{ 1 +\frac{\kappa^{2}}{\Omega_{R}^{2}}
\cos \Theta
\left[ 1 + \tilde{n} \right] \right\} \, ,
\label{growth rate}
\eea
where $\gamma=d_{b}^{2} |\kappav|^3/(3\pi\epsilon_0\hbar)$ is the free space
spontaneous emission rate, and $I_{d}$ is the intensity of the lasers.
Condensate growth corresponds to a positive rate $R$.  Assuming
that $\tilde{n} \gg 1$ this is equivalent to the following
requirement on the condensate phase difference
\beq
\cos \Theta <
- \frac{1}{\tilde{n}} \frac{\Omega_{R}^{2}}{\kappa^{2}}
 \, .
\eeq
This inequality is fulfilled by particular negative values of $\cos
\Theta$ provided that
$\tilde{n}$ is sufficiently large, and that the effective two-photon
detuning $2 \bar{\delta}$ is sufficiently small.  The latter may be
chosen small by manipulating the light-induced level shifts or the
relative frequency of the driving light beams.  The assumption of
equal wave numbers for the driving light fields in the calculations is
not very restrictive for the relative frequency, because an atom trap
introduces an uncertainty for the momentum conservation.  In the limit
of small two-photon detuning the effective linewidth of the transition
$c\rightarrow b$ may have an effect.  However, it may be shown to be
proportional to $\Delta_1^{-2}$ or smaller.

With a $1$ $\mu$m wavelength, 1 mW cm$^{-2}$ laser intensity, and
$(\kappa^{2}/\Omega_{R}^{2}) \cos \Theta = -1/2$,
the growth rate Eq.~{(\ref{growth rate})} is
\beq
R \approx N \tilde{n} \left( \frac{\gamma}{\Delta_{1}} \right)^{2}
  (10^{7} \mbox{\rm s}^{-1}) \, .
\eeq
A laser detuning of $\Delta_{1}/\gamma = 10^{3}$, condensates with
$N=10^{3}$, and $(8\pi/3) \tilde{n}=65$ give a condensate growth rate of $R
\approx 7 \times 10^{4}$ atoms per second.  This rate is large enough
to be useful for both atom laser pumping and for condensate growth.
However, sustained growth will require repopulation of the relevant
non-condensate atom modes by atom scattering processes.  Other
limitations on growth include diffusion of the condensate phase
difference due to atom-atom interactions \cite{Wong,WRI96}, and heating of
the non-condensate atom fraction by Raman transitions.  It is also
required that the spatial overlap of the condensate wave functions is
significant.

Multiple scattering can be ignored provided that only a small fraction 
of the incident photons are scattered.  If the cross sectional area of 
the condensate is ${\cal A}$, then this is true, if ${\cal A} \gg 6\pi 
(\gamma/\Delta_{1})^{2} N \tilde{n} / |\kappav|^{2}$.  With the 
preceding parameters this becomes ${\cal A} \gg 4 \times 10^{-3} \mu{\rm 
m}^{2}$, which is easily fulfilled.

In the calculations we used the conventional spontaneous symmetry 
breaking arguments that BECs are in coherent states.  However, this 
convenient approach is by no means necessary.  In fact, for the 
present atomic level scheme the relative phase between the two 
condensates has been established in stochastic simulations of the 
measurements of spontaneously scattered photons, even though the 
condensates are initially in pure number states \cite{Ruostekoski97}.  
Without any measurements of the condensate phase difference the 
macroscopic quantum coherence is expected to undergo collapses and 
revivals \cite{WRI96}.  Because the detections of spontaneously 
scattered photons establish the relative phase, they could also 
stabilize the phase against the collapse of the macroscopic wave 
function.

We have shown that quantum mechanical interference enables growth of
two Raman driven Bose condensates.  The growth mechanism is analogous
to that for gain in the laser without inversion. In that case
destructive quantum mechanical interference suppresses absorption,
allowing gain to dominate.  In our case interference suppresses
scattering of atoms out of the condensates, allowing scattering of
atoms into the condensates to dominate.

We would like to thank Howard Carmichael and Juha Javanainen for
useful comments.  This work was supported by the Marsden Fund of the
Royal Society of New Zealand, The University of Auckland Research
Fund, The New Zealand Lottery Grants Board, and The Australian
Research Council.

\end{document}